\documentclass[sn-mathphys]{sn-jnl}

\jyear{2021}%

\theoremstyle{thmstyleone}%
\theoremstyle{thmstyletwo}%

\theoremstyle{thmstylethree}%

\usepackage{acronym}
\usepackage{bm}
\usepackage[mathlines]{lineno}
\usepackage{amssymb}
\usepackage{amsfonts}
\usepackage{dsfont}
\usepackage{amsthm}
\usepackage{hyperref}
\usepackage{cleveref}
\usepackage{braket}

\newcommand{\ve}[1]{\ensuremath{\mathbf{#1}}} 
\newcommand{\RN}{\ensuremath{\mathbb{R}}} 
\newcommand{\indicator}{\ensuremath{\mathds{1}}} 
\newcommand{\slow}{SLOW} 
\newcommand{\fast}{FAST} 

\definecolor{lbColor}{rgb}{0,0.5,0}

\begin{document}

\title{Learning the noise fingerprint of quantum devices}


\author*[1,2]{\fnm{Stefano} \sur{Martina}}\email{stefano.martina@unifi.it}

\author[3,1]{\fnm{Lorenzo} \sur{Buffoni}}\email{lbuffoni@lx.it.pt}
\equalcont{These authors contributed equally to this work.}

\author[2,4]{\fnm{Stefano} \sur{Gherardini}}\email{gherardini@lens.unifi.it}
\equalcont{These authors contributed equally to this work.}

\author[1,2]{\fnm{Filippo} \sur{Caruso}}\email{filippo.caruso@unifi.it}

\affil*[1]{\orgdiv{Department of Physics and Astronomy}, \orgname{University of Florence}, \orgaddress{\street{Via Sansone, 1}, \city{Sesto Fiorentino}, \postcode{I-50019}, \state{FI}, \country{Italy}}}

\affil[2]{\orgdiv{European Laboratory for Non-Linear Spectroscopy (LENS)}, \orgname{University of Florence}, \orgaddress{\street{Via Nello Carrara 1}, \city{Sesto Fiorentino}, \postcode{I-50019}, \state{FI}, \country{Italy}}}

\affil[3]{\orgdiv{Physics of Information and Quantum Technologies Group}, \orgname{Instituto de Telecomunicaç\~oes, University of Lisbon}, \orgaddress{\street{Av. Rovisco Pais}, \city{Lisbon}, \postcode{P-1049-001}, \country{Portugal}}}

\affil[4]{\orgname{Scuola Internazionale Superiore di Studi Avanzati (SISSA)}, \orgaddress{\street{Via Bonomea, 265}, \city{Trieste}, \postcode{I-34136}, \state{TS}, \country{Italy}}}

\abstract{
Noise sources unavoidably affect any quantum technological device. Noise's main features are expected to strictly depend on the physical platform on which the quantum device is realized, in the form of a distinguishable fingerprint. Noise sources are also expected to evolve and change over time. Here, we first identify and then characterize experimentally the noise fingerprint of IBM cloud-available quantum computers, by resorting to machine learning techniques designed to classify noise distributions using time-ordered sequences of measured outcome probabilities.
}



\maketitle

\acrodef{nisq}[NISQ]{Noisy Intermediate Scale Quantum}
\acrodef{svm}[SVM]{Support Vector Machine}
\acrodef{ml}[ML]{Machine Learning}
\acrodef{svm}[SVM]{Support Vector Machine}
\acrodef{rbf}[RBF]{Radial Basis Function}
\acrodef{svc}[SVC]{Support Vector Classifier}
\acrodef{mmc}[MMC]{Maximal Margin Classifier}
\acrodef{sdk}[SDK]{Software Development Kit}

\section{Introduction}

In the quantum technologies context, no quantum device can be considered an isolated (ideal) quantum system. For this reason, the acronym \emph{Noisy Intermediate-Scale Quantum} (NISQ) technology has been recently introduced\,\cite{preskill2018quantum} to identify the class of early devices in which noise in quantum gates dramatically limits the size of circuits and algorithms that can be reliably performed\,\cite{deutsch2020harnessing,BhartiArxiv2021}. As early quantum devices become more widespread, a question that naturally arises is to understand, at the experimental level, whether in a generic quantum device the signature left by inner noise processes exhibits universal features or is characteristic of the specific quantum platform. Moreover, one may wonder to determine if such a noise signature has a time-dependent profile or can be effectively considered stable, in the sense of constant over time, while the device is operating. 

The answers to these questions are expected to be crucial in defining a proper strategy to mitigate the influence of noise and systematic errors\,\cite{DegenRMP2017,SzankowskiJPCM2017,DoNJP2019,MuellerPLA2020,wise2021using}, possibly going beyond standard quantum sensing techniques\,\cite{ColeNanotech2009,BylanderNatPhys2011,AlvarezPRL2011,YugePRL2011,Paz-SilvaPRL2014,NorrisPRL2016} and overcoming current limitations on probes dimension and resolution\,\cite{ColeNanotech2009,BylanderNatPhys2011,FreyNatComm2017,MuellerSciRep2018,HernandezPRB2018,GomezFrontiers2021}. On top of that, it gains even more importance in case one proves that noise signatures are peculiar to the single device, with the consequence that the issue of attenuating noise effects may be harder than expected. Indeed, each quantum technologies platform, ranging from superconducting circuits\,\cite{devoret2004superconducting,clarke2008superconducting} to trapped ions quantum computers\,\cite{wineland2003quantum}, photonic chips\,\cite{spring2013boson,metcalf2014quantum} and topological qubits\,\cite{freedman2003topological}, could need ad hoc solutions that usually are expensive and incompatible from a device to another. In addition, if the noise properties of a quantum device happen to be time-dependent, the system necessarily requires continuous calibrations, thus hindering not only the available runtimes, but also the accessibility from the external user and the replicability of the experiments performed on it. Furthermore, in case the noise fingerprint of the considered device can be easily discerned and remains unchanged over time, one could be able to identify from which specific quantum device certain data were generated just by looking at the noise fingerprint. However, this aspect might create problems, in principle, for possible future usages of the device in privacy-sensitive applications.

In this paper, we aim to shed light on the previously discussed aspects by providing a powerful tool, based on \ac{ml} techniques, for the classification of noise fingerprints in quantum devices with same technical specifications but physically placed in different environmental conditions. \ac{ml}\,\cite{BishopPRML2006, HastieESL2009} -- originally introduced in the classical domain to learn from data, identify distinctive patterns, and then make decisions with minimal human intervention -- has been already proven useful to characterize open quantum dynamics\,\cite{YoussryArXiv,luchnikov2020machine,fanchini2020estimating} and to carry out quantum sensing tasks\,\cite{niu2019learning,HarperNatPhys2020,MartinaArXiv2021,wise2021using}, as for example the learning and classification of non-Markovian noise\,\cite{niu2019learning,MartinaArXiv2021} or the detection of qubits correlations\,\cite{HarperNatPhys2020}.

Here, we first design a quantum circuit that mimics the transport dynamics of a quantum particle on a network of $16$ nodes that are identified by the states of the computational basis 
$\{\lvert 0000\rangle,\lvert 0001\rangle,\dots,\lvert 1110\rangle,\lvert 1111\rangle\}$. The designed quantum circuit is measured (by locally applying the Z Pauli operators on some qubits of the circuit) in $9$ distinct parts that, from now on, we denote as \emph{measurement steps}. The routine allowing to record all the outcome in each measurement step is instead denoted as \emph{execution}. Moreover, the repetition of a given number of executions is called \emph{run}. Employing the open-access quantum computers offered by the IBM Quantum Experience\,\cite{ibmq}, we experimentally classify a set of quantum devices by executing in all of them the same quantum transport circuit (testbed circuit). The classification is enabled by the presence of a peculiar noise fingerprint associated to each quantum machine. In more details, the \ac{ml} models are trained by taking as input the distributions of the outcomes recorded at the $9$ measurement steps of the testbed circuit. As shown in the next sections, the classification is successfully achieved with a test accuracy greater than $99\%$, both on diverse IBM machines and on single devices but at different times from one execution to another. Indeed, from our experiments we can observe that the noise fingerprint of each tested quantum devices has also a clear time-dependence, meaning that executions of a quantum circuit, implemented at different times, can be associated to distinctive main traits. 

These experimental evidences lead us to the conclusion that different IBM quantum devices exhibit distinctive, and thus distinguishable, noise fingerprints that, however, can be characterized and predicted by \ac{ml} methods. Therefore, the proposed solution might be pivotal to certify the time-scheduling and the specific machine on which a given quantum computation is executed. Moreover, learning the noise fingerprint of the quantum device under analysis could play a key role both for diagnostics purposes -- especially in all those contexts where logic quantum operations cannot be error-corrected\,\cite{deutsch2020harnessing} and thus need to be ``noise-free'' as much as possible -- and to accomplish benchmarking and certification\,\cite{eisert2020quantum} of quantum noise sources within a pre-established error threshold. 

\section{Experimental platform}

For our experiments we employ the IBM Quantum cloud services to run remotely quantum circuits on several machines. To interact with the remote services, we use the Qiskit \ac{sdk}\,\cite{qiskit}, which is an open-source Python \ac{sdk}, useful both to \textit{simulate} quantum dynamics (with or without noise) and to \textit{program} a given set of operations on a real quantum device. Overall, we have at our disposal up to $11$ superconducting quantum computers ranging from a single qubit up to $15$ qubits, with different topology and calibration routines. For all the available devices and their characteristics, we direct the reader to the IBM documentation\,\cite{ibmq}.

The accessibility and availability of the IBM devices allow to carry real experiments having the flexibility of taking either a lot of samples in a short amount of time, or collecting samples from the same circuit but at longer time intervals. As it will be shown below, both these aspects will be properly exploited in carrying out our experiments. Moreover, one can also run the same exact circuit not only on a single device but on multiple machines, thus enabling the creation of complete datasets of quantum experiments to be fed in \ac{ml} algorithms. Regarding the generation of our datasets, we refer the reader to the source codes at the address provided at the end of the manuscript. 

Overall, several experiments (explained in detail later) have been conducted on different IBM chips (specifically, `Yorktown', `Athens', `Bogota', `Casablanca', `Lima', `Quito', `Santiago', `Belem', and `Rome'). The chips differ by two main aspects. The first is the architecture (or connectivity) of the qubits, which ranges from a simple line topology to a ladder or a star topology. The second important difference is the so-called quantum volume\,\cite{cross2019validating} ($8$, $16$, $32$ for the machines used in our experiments) that quantifies the maximum dimension of a circuit that can be effectively executed, and is correlated also with the noise affecting each device. Indeed, some quantum machines are inherently noisier than other, and even single qubits inside a machine can have a distinctive noise profile. All these peculiar differences in noise and topology represent the fingerprint that we aim to exploit using our method. 

Before proceeding, it is worth stressing that, albeit the proposed experiments are carried out on superconducting devices, the gate-model approach adopted here is valid in principle for a large class of NISQ devices.

\section{Testbed quantum circuit}

\begin{figure}[t!]
\centering
\includegraphics[width=0.9\linewidth]{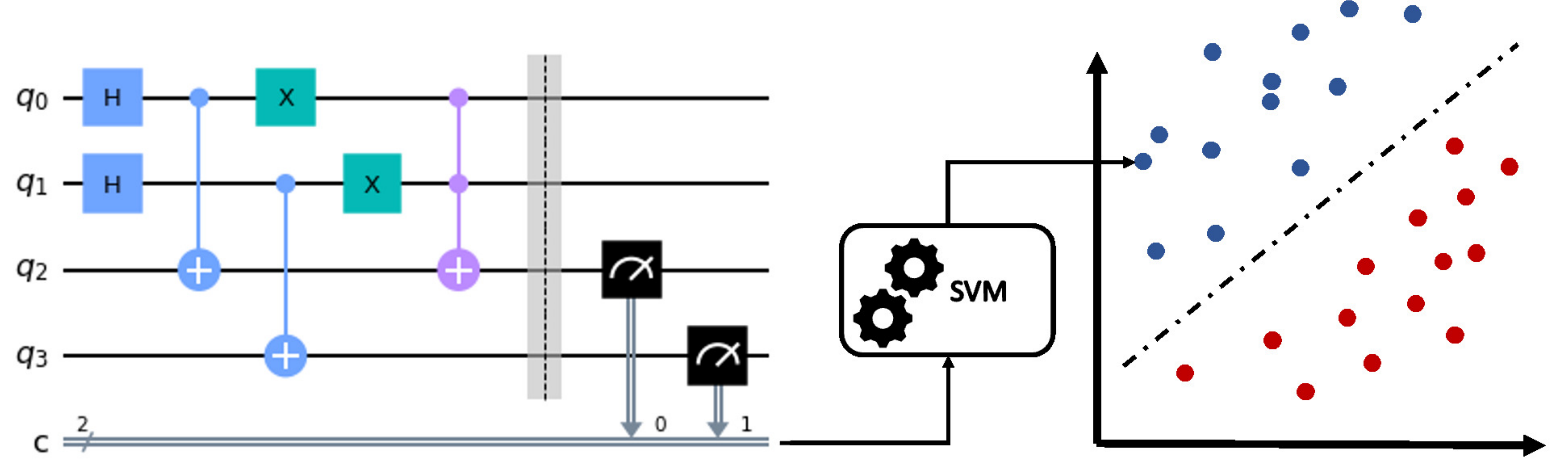}     
\caption{On the left, circuit implementation of the quantum transport dynamics employed as a testbed. The quantum circuit, which involves 4 qubits, is repeated more than once (in our experiments, $3$ times), and 2 of the 4 qubits are measured at regular steps. The outcome probabilities obtained by our measurements, which together form the datasets to train, validate and test the used \ac{ml} models, are fed into a Support Vector Machine (SVM) -- schematically represented on the right -- in order to be classified.}
\label{fig:quantum_circuit}
\end{figure}

To learn the noise fingerprint of IBM quantum devices, we employ a quantum walker on a network of $16$ nodes realized by the quantum states $\lvert 0000\rangle,\lvert 0001\rangle,\dots,\lvert 1110\rangle,\lvert 1111\rangle$ through the circuit in Fig.\,\ref{fig:quantum_circuit}. Notice that, for our purposes, the number of qubits of the testbed circuit can be just a few; however, this does not imply that the proposed solutions cannot be applied to circuits with generic dimension.

The idea of our testbed circuit is to simulate a quantum transport dynamics, whereby, by initialising a quantum particle in one node of the network (specifically, in the state $\lvert 0000\rangle$), the particle ``flows'' across the allowed pathways thanks to the action of local operations and of \emph{controlled NOT} (CNOTs) and \emph{Toffoli} gates (denoted in Fig.\,\ref{fig:quantum_circuit} by a light blue and light purple rounds, respectively, with the symbol `plus' inside). We recall that the CNOT is a two-qubit quantum operation, commonly used to entangle/disentangle Bell states, that flips the second qubit when the first qubit is in $\lvert 1\rangle$. Instead, the Toffoli gate is a universal ``controlled-controlled-not'' (3-qubit) operation where a third qubit is flipped when two control qubits are both in $\lvert 1\rangle$. In our circuit in Fig.\,\ref{fig:quantum_circuit}, two qubits (i.e., $q_3$ and $q_2$ in the figure) are used to get information on the particle, providing at each measurement the pair of bits $(0,0),(0,1),(1,0),(1,1)$, where the first and second bits correspond, respectively, to the outcomes measured on $q_3$ and $q_2$. Conversely, qubits $q_0$ and $q_1$ are employed as ancilla qubits to move, and thus control, the particle. 
Then, this quantum circuit is repeated $3$ times, with the aim to collect data on the quantum dynamics in each IBM device. As already mentioned in the Introduction, the resulting quantum circuit (given by repeating $3$ times the circuit in Fig.\,\ref{fig:quantum_circuit}) is locally measured in $9$ distinct parts (corresponding to the measurement steps) thanks to the simultaneous application of Z Pauli operators $\sigma_z$ on the qubits $q_3$ and $q_2$, from which the measurement outcomes are collected. It is worth noting that the procedure we are proposing is not based on repeated measurements as in a quantum monitoring protocol or in Zeno quantum dynamics \cite{FischerPRL2001,SchaferNatComm2014,GherardiniQST2017,VirziArXiv2021}, since, each time a measurement is performed at a given measurement step (say the $k$-th, with $k=1,\ldots,9$), the whole testbed quantum circuit is regenerated and then (locally) measured at the subsequent step, i.e., the $(k+1)$-th.

In a single repetition, the quantum circuit is initialized in $\lvert 0000\rangle$ that corresponds to the measurement outcomes $(0,0)$, and then two Hadamard gates (blue squares `H' in Fig.\,\ref{fig:quantum_circuit}) are applied to both $q_0$ and $q_1$. Thus, since the two CNOT gates are conditioned to $q_0$ and $q_1$ respectively, the probability to get $1$ or $0$ in $q_2$ and $q_3$ after the CNOTs is $0.5$. In this way, after the Pauli-X rotation (green squares `X' in Fig.\,\ref{fig:quantum_circuit}) and the Toffoli gate, the system is in the state $\frac{1}{2}(\lvert 0110\rangle+\lvert 0111\rangle+\lvert 1001\rangle+\lvert 1100\rangle)$ before that the qubits $q_2$, $q_3$ are measured along the $z$-axis (black squares in Fig.\,\ref{fig:quantum_circuit}). This entails that, at the end of the circuit, measuring $q_3$ and $q_2$ provides the results $\{(0,0),(0,1),(1,0),(1,1)\}$ with probabilities respectively $\{0,0.5,0.25,0.25\}$. Of course, such a dynamic only occurs under \emph{ideal} unitary evolution, which is not the case of the implementation on real experimental devices. In our case, the noisy environment, in which the machines are immersed, alters each realization of the simulated quantum (transport) dynamics, thus making stochastic the evolution of the particle within the circuit. As we will prove below, this randomness is a specific feature of each machine and changes from one device to another, thus allowing us to perform classification tasks. Specifically, are the discrepancies between the measured outcome probabilities (from qubits $q_2$ and $q_3$) on one or more IBM machines that enable to learn the corresponding noise fingerprint, and then classify from which device the input data have been generated. Here, it is worth noting that, despite from one implementation to another a slight different physical Hamiltonian may be implemented in the chips of each quantum device, the variations observed in the measurement outcome distributions -- having a prominent random nature -- are not ascribable to such a deterministic aspect, but to a stochastic cause thus pertaining to an external noise source. However, a same stochastic process can affect differently two equivalent quantum dynamics but originated by two distinct physical Hamiltonian operators. Therefore, the fingerprint that we leverage for the classification can be due not only to differences in the noise profiles affecting the quantum devices, but also on their dependence on the way the testbed circuit is physically implemented.

While our picture of considering the implemented quantum dynamics as the ones of a quantum walker may be pretty useful for illustrative purposes, we want to stress that this kind of dynamics has been chosen just for its simplicity and generalizability with other types of NISQ devices. Indeed, our results shown in the following are quite general, since they do not depend on specific dynamics and do not require initial assumptions. Accordingly, we expect that such results may be re-obtained in other quantum devices, even ones not necessarily designed to carry out computing tasks.

\section{Machine learning model}

Let us provide some details on the adopted \ac{ml} model, i.e., the popular \ac{svm}\,\cite{HastieESL2009}. 

The dataset yielded as input to the \ac{svm} is a set of $n$ points $\ve{x}_q\in\RN^p$, with $q=1,\ldots,n$, each of them living in the $p$-dimension space of the data features, where a feature is a distinctive attribute of the data set elements. 

In binary classification problems, to each $\ve{x}_q$ with $q=1,\dots,n$ is associated a class $y_q\in\{-1,1\}$ that represents the desired output of the \ac{svm}. By contextualizing it to our problem, the binary classes $y_q$ denote if a given set of points $\ve{x}_q$ have been generated $(+1)$ or not $(-1)$ by a specific machine or in a time window/interval. 
A \ac{svm} for binary classification is trained such that the two classes of points (provided as input to the \ac{ml} model) are separated by the hyperplane that maximises the distance between the hyperplane itself and the nearest points of the classes (commonly denoted as \emph{margin}). If the points $\ve{x}_q$ of the data set are not linearly separable (which is most often the case), then the value of the margin is negative and the points cannot be classified. To circumvent this problem, \acp{svm} employ a clever mapping in an higher-dimensional space (called feature-space) with polynomial or \ac{rbf} kernels that allows for an easy classification as in Fig.\ref{fig:quantum_circuit}. The extension to multiclass classification problems is then obtained by associating a class with multiple values to each $\ve{x}_q$. In our experiments, part of the generated dataset is used as a validation set to choose the best mapping among the kernels: linear (meaning that the data is already linearly separable), polynomial with degree 2, 3 and 4, and \ac{rbf}. In many cases, just the simple linear kernel is enough to successfully perform the classification, but in other cases (e.g., in multiclass classification) the more complex kernels may be beneficial.

Finally, in our experiments, the classification \emph{accuracy} is computed by comparing the predictions $\hat{\ve{y}}$ returned by the \ac{ml} models with the desired classes $\ve{y}$ of the test set:
\begin{equation}
    {\rm accuracy}(\hat{\ve{y}}, \ve{y})\equiv\frac{1}{n}\sum_{i=1}^n\indicator\left\{\hat{\ve{y}}_i = \ve{y}_i\right\}
\end{equation}
where $\indicator\{\cdot\}$ is the \emph{indicator function}
such that $\indicator\{c\}=1$ if $c$ is true, and $\indicator\{c\}=0$ otherwise. In this regard, to clarify the naming convention for the reader, we refer to: ``training'', ``validation'' and ``test'' sets, to identify three non-overlapping partitions of the data. These partitions are used respectively to: train the model, validate the best parameters, and test the performance on unseen data. In the experiments we randomly select $60\%$ of the data to train the \ac{svm} model, $20\%$ to validate different configurations (i.e., \ac{svm} kernel type), and $20\%$ to report the results on unseen data.

\section{Experiments description}\label{sec:experimentsDescription}

The results, which we are going to show, concern three series of \ac{ml} experiments that use two different datasets, obtained from the IBM quantum chips mentioned above.

In the first two experiment series, the \ac{ml} models are trained both to discriminate the noise fingerprint of different quantum devices and to identify a time-dependence in each of them. The training of some of the models is performed on the dataset here denoted as \fast{} that collects the outcome distributions measured in temporally-close executions of the testbed quantum circuit on $7$ different IBM quantum machines (i.e., `Athens', `Bogota', `Casablanca', `Lima', `Quito', `Santiago', `Yorktown'). In these experiments, $20$ parallel tasks (corresponding to the maximum allowed number) are appended to the IBM fair-share queue, and, once a task is concluded, another task is immediately added. For each task the testbed circuit is run $8\,000$ times for each one of the $9$ different steps, and the probabilities to get the measurement outcomes are computed over $1\,000$ shots among the total $8\,000$ to obtain $8$ different outcome probabilities times $9$ steps per task. 

\begin{figure}[t!]
\centering
\includegraphics[width=0.7\linewidth]{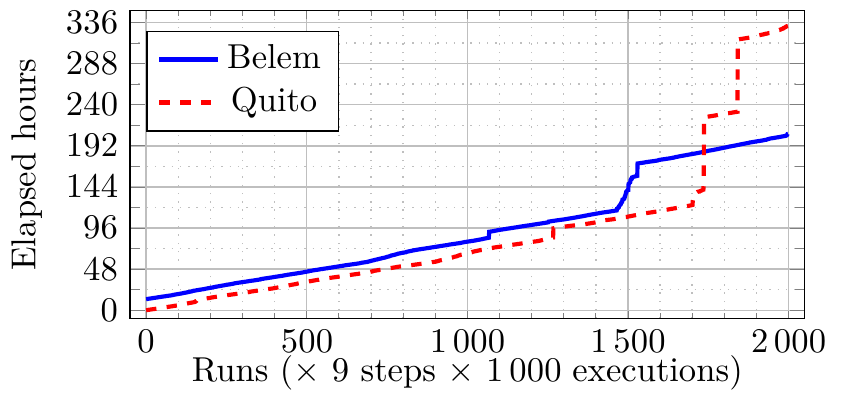}     
    \caption{
    Elapsed hours to collect all the measurement outcomes on the IBM machines `Belem' and `Quito' (solid blue and dashed red lines, respectively) for the dataset \slow{}. Each point of the curves, obtained over $1\,000$ executions of the testbed quantum circuit
    for each measurement step $k=1,\ldots,9$, is associated to the relative physical/real time in which the measurement probabilities are computed in a single run. Notice that, if compared with the time scale of the vertical axis (y-axis), which is expressed in hours, the computation of the $9\,000$ executions of each run can be considered practically instantaneous, i.e., in the order of some seconds. Moreover, the anomalous behaviour of the curves after $1\,500$ runs has to be attributed to the policy of the IBM fair-share queue.
    }
    \label{fig:walkerSlow}
\end{figure}

Conversely, in the third \ac{ml} experiment series, we perform a robustness analysis by making stricter the time constraints on the employed datasets. Specifically, in those experiments, and in part of the previous ones, we employ a second dataset, called \slow{}, which is composed of measurement distributions extracted from executions in two different quantum machines (`Belem' and `Quito') more ``slowly'' than the data in the first dataset. As represented in Fig.\,\ref{fig:walkerSlow}, more ``slowly'' means that only one task per time is appended to the queue and then run, waiting at least $2$ minutes from the conclusion of the previous task. Moreover, for each task the testbed circuit is executed, for each one of the $9$ steps, $1\,000$ times that corresponds to the number of shots set to compute the outcome distributions.

We recall that in each execution, for both the \fast{} and \slow{} datasets, the qubits $q_3$, $q_2$ of the testbed quantum circuit (the full circuit is obtained by repeating 3 times the circuit in Fig.\,\ref{fig:quantum_circuit}) are measured iteratively after each CNOT and Toffoli gate, for a total of $9$ outcome probabilities at the consecutive measurement steps $k = 1,\dots,9$. Overall, for each machine, we have collected $2\,000$ sequences of $9$ probability distributions built with the measurement outcomes from the qubits $q_3$ and $q_2$ of the testbed circuit. This means that a total of $2\,000\,000 \times 9$ single executions have been run on each quantum machine that we employed to generate the \fast{} dataset, and similarly for the \slow{} one.

As final remark, let us note that the \fast{} dataset is employed for the experiments illustrated in section \ref{sec:quantumDevicesClassification} and part of section \ref{sec:noiseFingerprintAtDifferentTimeScales}, while the \slow{} dataset to complete the experiments in section \ref{sec:noiseFingerprintAtDifferentTimeScales} and perform in \cref{sec:robustnessAnalysis} a robustness analysis at different time scales.

\section{Quantum devices classification}
\label{sec:quantumDevicesClassification}

As first, we present binary classification experiments. For each pair of IBM machines, a \ac{svm} model is trained using the dataset \fast{} (introduced in \cref{sec:experimentsDescription}) with the aim to identify on which device the executions of the testbed quantum circuit are run.   
The inputs of the \ac{svm} model are the distributions of the measurement outcomes from qubits $q_3$ and $q_2$ recorded at the discrete measurement steps $k=1,\dots,9$. Specifically, two different kinds of inputs are set: In the first we consider only the outcome distributions measured at the single step $k$ with $k\in[1,9]$, while in the second we concatenate all the measurement probabilities in ordered sequences $1,\dots,k$. Then, our \ac{ml} experiments are performed by alternatively taking the two types of inputs; we will report below the resulting accuracy values for both of them.

\begin{table}[]
    \centering
    \caption{Classification accuracy, denoted as $\alpha({\cdot})$, of all the possible binary \ac{svm}s trained with the measurement probabilities collected in the dataset \fast{}. For each experiment, a large number of executions are run on two different IBM machines (whose names are in the $1$st column and in the $1$st row of the table) in correspondence of the measurement steps $k$ ($1$st column of each sub-table). Then, two different inputs are tested: Outcome distributions at single steps (whose accuracy values are in the $2$nd column of the sub-tables) and sequences of measurement probabilities obtained at each $k$ (accuracy values in the $3$rd column of each sub-table).
    \\}
    \includegraphics[width=\linewidth]{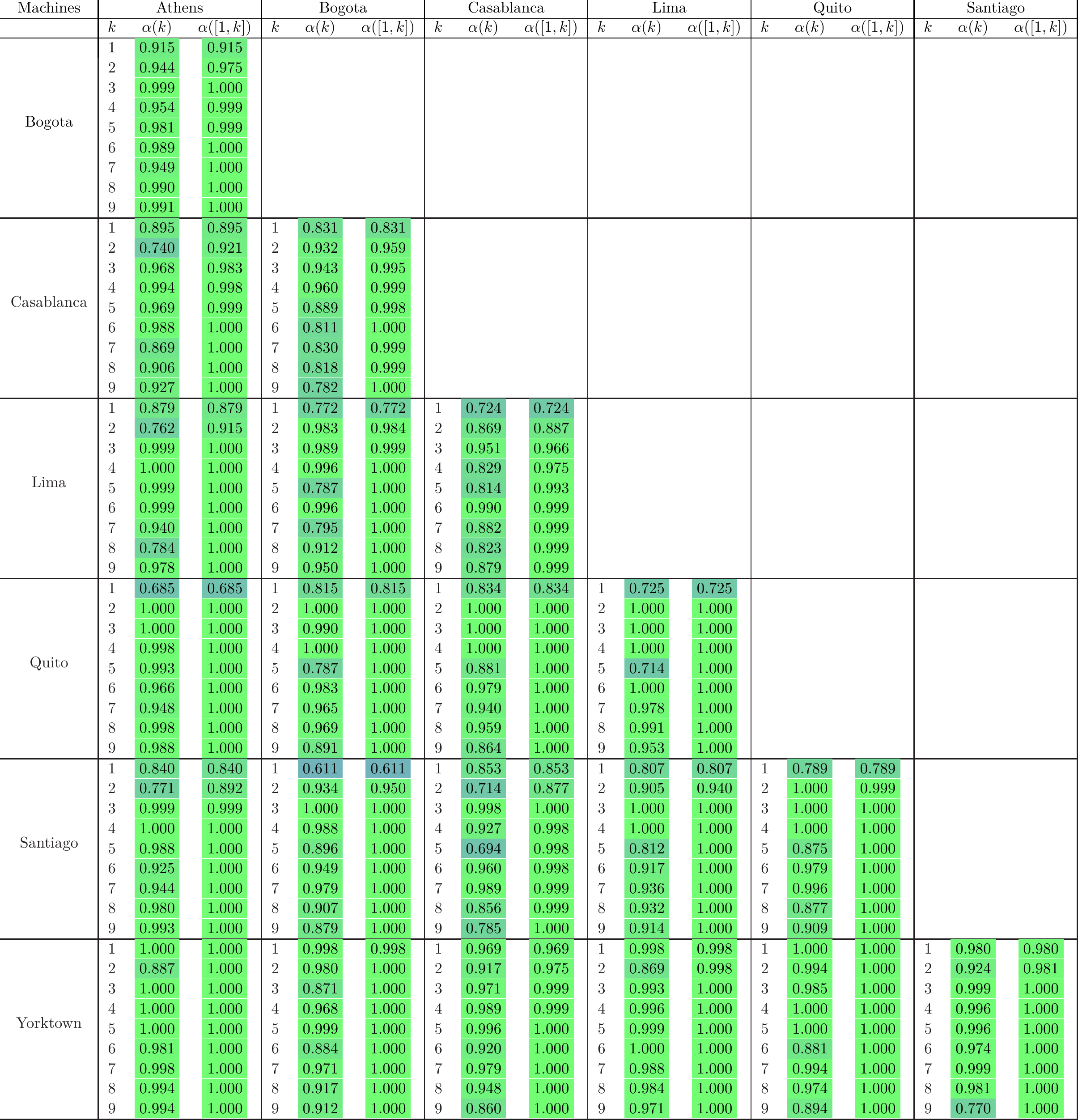}
    \label{tab:allResults}
\end{table}
From the results of our experiments -- reported in \cref{tab:allResults} -- we observe that it is sufficient to use only the outcome probabilities corresponding to the first three measurements at $k=1,2,3$ to reach more than $99\%$ of accuracy in discriminating all the pairs of tested machines. This implies that, in a realistic deployment scenario, one needs less data than the amount acquired here to reach good classification performances. An additional observation we can make is that the accuracy is \emph{not monotonic} in $k$ when considering the classifier using single measurement data. This can be due to the fact that, at various measurement steps, to distinguish the noise fingerprint from a single measurement probability might be easier or harder. On the other hand, we can also observe that the accuracy is steadily increasing when as input is set the sequence of all outcome distributions up to any measurement step $k$. Hence, from this we can deduce that, to identify the noise fingerprint of IBM quantum devices, sequences of outcome distributions recorded at more than one measurement steps need to be taken into account.  
This is also the reason why we deem important to frame the issues addressed in this paper as belonging to a noise fingerprint \emph{in time} instead of single shot measures. 

\begin{table}[t!]
    \centering
    \caption{Classification accuracy, denoted as $\alpha({\cdot})$, of multiclass \ac{svm}s trained with the measurement probabilities collected in the dataset \fast{}. A large number of executions are run on 7 different IBM machines ($1$st column of the table) in correspondence of the measurement steps $k\in[1,9]$ ($2$nd column). Different inputs are tested: Outcome distributions at single steps ($3$rd column), sequences of measurement probabilities computed on windows of width from $2$ to $5$ steps before each $k$ (from $4$th to $7$th columns), and the sequences of all measurement probabilities obtained from the $1$st to the $k$th step ($8$th column). Finally, the last row of the table reports, for all the $k$s, the averages of the accuracy values in the 
    rows above; the average of the last column is omitted since the accuracy values therein are calculated on models with different numbers of input measurement steps.\\}
    \includegraphics[width=\linewidth]{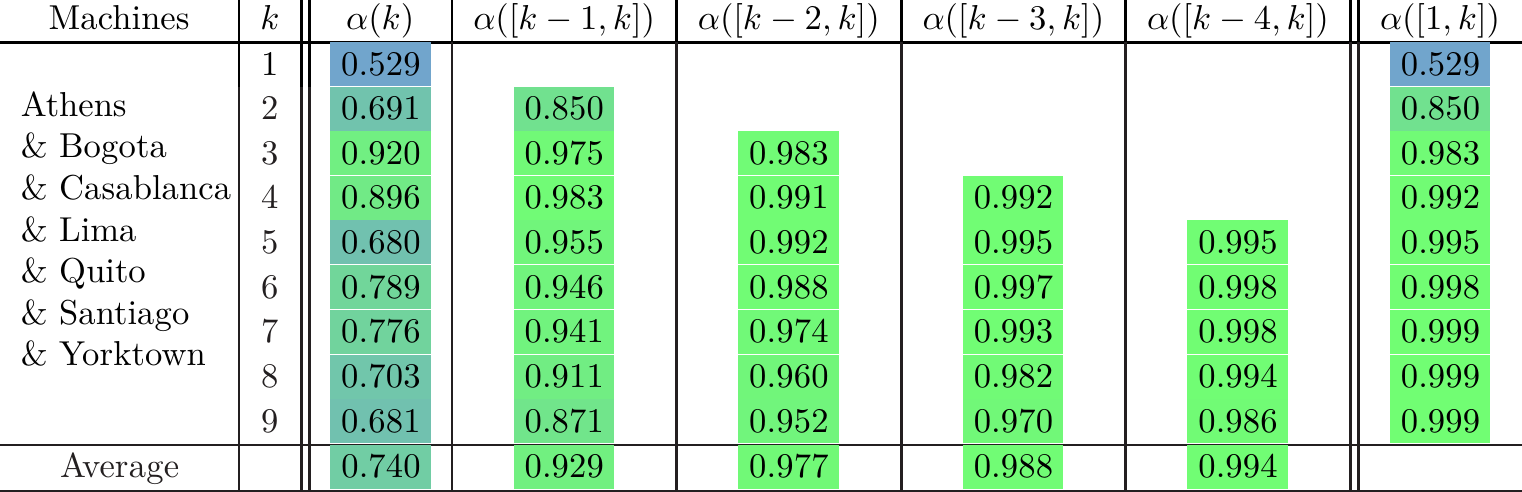}
    \label{tab:multiclassClassification}
\end{table}

Let us now extend the binary \ac{svm} algorithms to multiclass classification problems, in which more quantum devices are simultaneously discriminated. In our experiments, the so-called \emph{one-vs-rest} strategy is adopted\,\cite{BishopPRML2006}, where for $n$ distinct classes we train $n$ different binary classifiers that discriminate the elements of a class from the others. In particular, our multiclass \ac{svm} is trained with the aim to identify to which IBM quantum machines, among the $7$ that have been used, belongs a given set of measured outcome probabilities (from the testbed quantum circuit) of the \fast{} dataset. The results in \cref{tab:multiclassClassification} reports the test accuracy values returned by the models that are trained with different input data. As in binary classification, for one kind of input data, the model is trained with the outcome distributions obtained at single step $k$ with $k\in[1,9]$ ($3$rd column of \cref{tab:multiclassClassification}), while another set of input data is provided by concatenated measurement probabilities $1,\dots,k$ ($8$th column). 
Moreover, for the purpose of multiclass classification, 
further input are also adopted: At each step $k$ the model is trained not only with the outcome distributions at the $k$th step, but also with a window of preceding measurement probabilities belonging to $[k-s,k]$ with $s$ integer number. Regarding $s$, the range from $1$ ($4$th column of \cref{tab:multiclassClassification}) to $5$ ($7$th column of \cref{tab:multiclassClassification}) is considered.
As for the binary case, the \ac{svm} is able to successfully discriminate between the tested machines just by using the measurement outcomes taken in few measurement steps. While the accuracy using the outcomes at single-time measurement steps oscillates, the time-ordered sequence monotonically increases. That confirms our previous observations about the need of a time sequence to have a reliable fingerprint. In addition, the models trained with the input data on sliding windows allow us to understand the effective need of outcome distributions taken from more than a single measurement step for the classification of the noise fingerprint. In such case, we observe that the accuracy at each step $k$ steadily increases with the size of the set of considered steps, and this holds also by looking at the average of the accuracy values computed over all the measurement steps. It is worth noting that the last column on \cref{tab:multiclassClassification} expresses a similar strategy, where the single accuracy values are provided as output of the models trained on a window (with increasing dimension) that always starts from the $1$th to the $k$th step. In other words, the first accuracy values on top of columns from $3$ to $7$ correspond to the elements of the last columns for $k$ from $1$ to $5$.

The high-level of accuracy (even more than $99$\%) in carrying out binary and multiclass classification of the IBM quantum machines is an evidence for the presence of a strong underlying noise fingerprint in the dynamics of NISQ devices. Indeed, this is the key feature that can allow one to identify, basically in a deterministic way, from which quantum machine a given set of measurement has been obtained. 

\section{Noise fingerprint at different time scales}
\label{sec:noiseFingerprintAtDifferentTimeScales}

Since the environment of the IBM quantum devices changes quite often (e.g., the machines are calibrated up to multiple times in an hour), we have slightly modified our experiments to prove also the existence of a noise fingerprint that pertains to the temporal evolution of the chip on which a given quantum circuit is executed. 
\begin{table}[t!]
    \centering
    \caption{Classification accuracy, denoted as $\alpha(\cdot)$, of \acp{svm} -- trained with two sets of outcome distributions from the dataset \fast{}, temporally separated by 24 hours -- to predict in the IBM machine `Casablanca' which executions were implemented the first day and which the second day. Also in this case, the inputs to the \ac{svm}s at the measurement steps $k$ ($2$nd column of the table) are the outcome distributions at single steps ($3$rd column) or the sequences of measurement probabilities computed at each $k$ ($4$th column).\\}
    \includegraphics[width=0.45\linewidth]{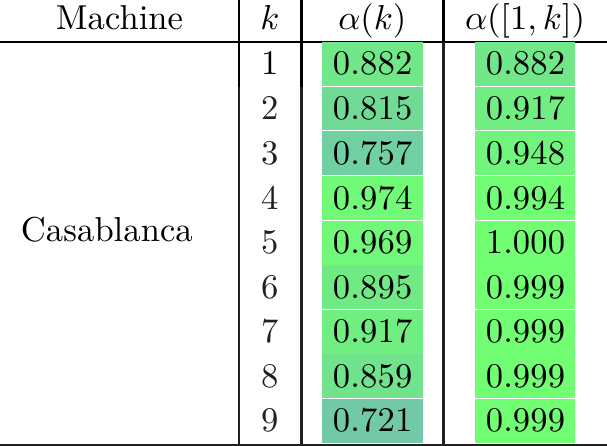}
    \label{tab:binaryTemporalClassification}
\end{table}
To confirm this hypothesis, we have designed a temporal classification setting that we employ with data from both the \fast{} and \slow{} datasets.

Regarding the experiments using the \fast{} dataset, two sets of measurement outcome distributions are collected for the machine `Casablanca', one temporally separated from the other by 24 hours. After that, similarly to what done in the previous experiments, a \ac{svm} model is used to discriminate the executions implemented the first day on the IBM device from the ones performed on the second day. From these experiments, whose results are shown in \cref{tab:binaryTemporalClassification}, we observe that the designed \ac{ml} algorithms are able to detect a characteristic fingerprint, still induced by the presence of noise sources, in a single quantum device but in measurement steps separated by a quite long ($24$ hours) time interval. In such classification tasks, an accuracy of $95\%$ is achieved by the \ac{ml} models, just by taking as input the sequence of outcome distributions at the first measurement steps $k=1,2,3$. 

\begin{table}[t!]
    \centering
    \caption{Binary classification accuracy, denoted as $\alpha(\cdot)$, of \acp{svm} trained to classify the outcome distributions belonging to 
    distinct two sets of data. One set is composed by the runs of `Belem' in the \slow{} dataset by numbering them from $1$ to $200$ in temporal ordering. Also the other set is composed by runs of `Belem' in the \slow{} dataset, but collected within temporal windows specified on the columns title (from run $201$ to run $400$, from run $401$ to run $600$, etc\dots). In the top sub-table the models are trained with the outcome distributions taken at the $k$th measurement step, while in the bottom sub-table the inputs are the sequences of measurement probabilities from step $1$ to step $k$.
    \\}
    \includegraphics[width=\linewidth]{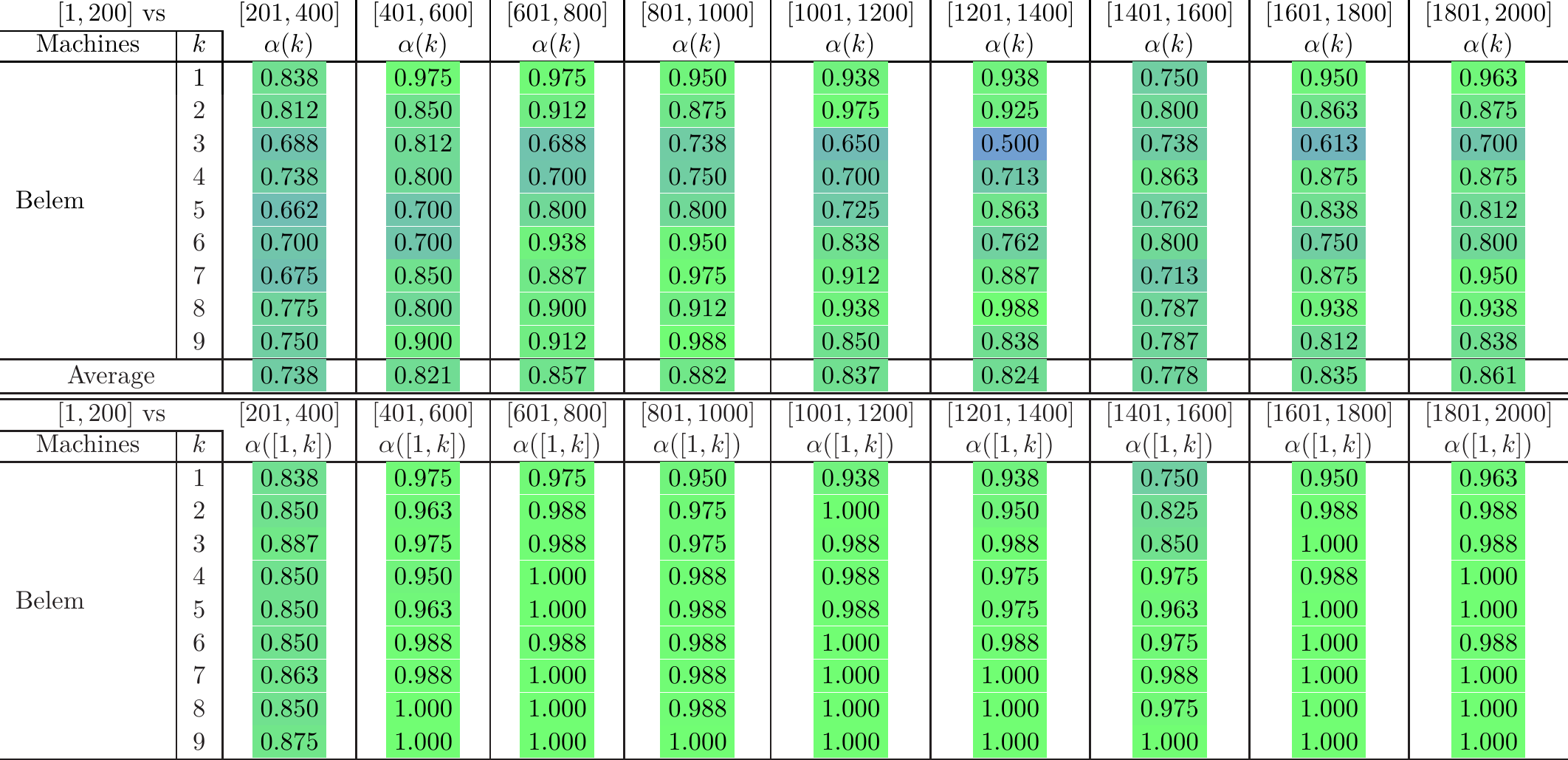}
    \label{tab:windowTemporalClassification}
\end{table}
In order to better quantify the evolution in time of the noise fingerprint, we use data from the runs of `Belem' in the \slow{} dataset. Respect to the previous dataset, the data from the runs in \slow{} are more evenly distributed in time so that we have decided to split the data in $10$ adjacent windows, each of them containing $200$ consecutive runs. Subsequently, the \acp{svm} models are trained to classify if a run has been computed on the first window (from run $1$ to run $200$) or in another window of the remaining $9$. From the results in \cref{tab:windowTemporalClassification}, we can observe that is difficult to distinguish the runs pertaining to the first window from the runs in the adjacent window (i.e., runs from $201$ to $400$ in the third column), either considering as input the single outcome distributions at the $k$th measurement step (the top part of \cref{tab:windowTemporalClassification}) or the sequences of measurement probabilities from step $1$ to step $k$ (bottom part). As a matter of fact, we do not reach $90\%$ in neither case. Conversely, when we consider the subsequent windows (runs after $400$ on the next columns), thus at a greater distance from the first window, the classification task becomes easier. 

Analogously to the previous experiments, the single measurement outcomes do not seem to carry enough information on the noise fingerprint and the classification accuracy depends on the choice of $k$. Instead, when we consider the sequences of outcomes for all the steps, we can observe that 
the noise fingerprint in the first window of runs can be much better distinguished from the corresponding fingerprint in all the subsequent windows, except the neighbouring one. The window from run $1401$ to run $1600$ seems more challenging to classify with respect to the others. One possible reason for this can be that, as one can see from \cref{fig:walkerSlow}, around the run $1500$ the policies of the IBM fairshare queue caused a discontinuity in time. This means that the data distribution inside the aforementioned window has more variance with respect to the data in the other windows and for the \ac{ml} models can be find more difficult to classify the data. However, even in that case the classification accuracy reaches $100\%$ when using the sequence of measurement probabilities for all the steps $k=1,\dots,9$.

\begin{figure}[t!]
\centering
\includegraphics[width=0.7\linewidth]{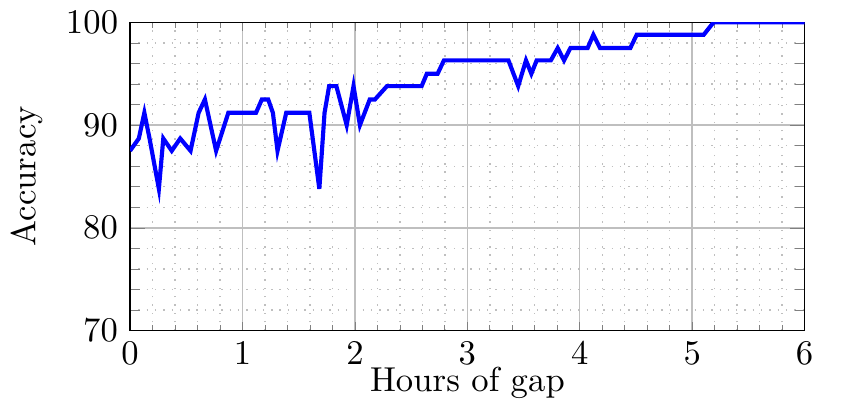}     
    \caption{
    Maximum reached accuracy for \ac{svm} models trained on sequences of measurement outcomes for all the steps $k=1,\dots,9$ taken from the `Belem' quantum machine and collected in the dataset \slow{}. The model is trained to classify the executions in the window of runs from $1$ to $200$ from the ones in a subsequent window of $200$ runs. Initially, the latter is adjacent to the first window, then it is moved by increasing the gap between the two windows. The plotted curve is then obtained by drawing the accuracy values for the corresponding gaps, expressed in hours. Note that a gap of $6$ hours correspond to approximately $90$ runs.
    }
    \label{fig:timeAccSlidingWindow}
\end{figure}
In these experiment, the execution times for all the runs in each window is approximately $12$ hours (except for the previously-discussed window from run $1400$ to $1600$). Thus, we can deduce that $12$ hours of time distance between the windows are sufficient to distinguish the noise fingerprint at different times with $100\%$ of accuracy. To find the minimum necessary hours gap, in \cref{fig:timeAccSlidingWindow} we report the reached accuracy of a \ac{svm} model trained to distinguish the runs in the first window (from run $0$ to $200$) of `Belem' within the \slow{} dataset from the runs in another window with an increasing time gap among them. We can observe that, in this case, already after $6$ hours the noise fingerprint is distinguishable with an accuracy of $100\%$. In general, we can observe that even starting from different windows in time, and using different window sizes, more than $95\%$ of accuracy is reached after a few hours (in the order of one day).

Overall, we can thus conclude that a clear temporal dependence of noise fingerprint is present in our experiments, even when the same quantum machine is taken into account.

\section{Robustness analysis}
\label{sec:robustnessAnalysis}
Finally, we investigate the robustness of the learned fingerprint at different time scales. For this purpose, taking the IBM machines `Belem' and `Quito', we temporally order all the executions of the testbed quantum circuit, by dividing them in $10$ distinct windows of $400$ consecutive runs, i.e., $200$ runs per machine. The elapsed time between runs has been already reported in Fig.\,\ref{fig:walkerSlow}. In this way, after have generated the \slow{} dataset (introduced in \cref{sec:experimentsDescription}) with $2000$ runs per machine, the \ac{svm}s are trained to classify 
on which device, among `Belem' or `Quito', the testbed quantum circuit has been executed.
Specifically, in any experiment designed for the robustness analysis, the \ac{ml} model is trained over the data collected in a time window of $200$ consecutive runs (overall, we consider $10$ distinct time windows), and then tested in all the considered time windows including the one used for the training.

\begin{table*}[!ht]
\caption{Classification accuracy of \ac{svm}s trained to classify 
on which quantum device, among `Belem' or `Quito', a given set of data has been generated. The training of the models is performed with the outcome distributions collected in the dataset \slow{}, and then divided in $10$ distinct time windows of $200$ runs (the first window includes the runs from $1$ to $200$, the second from $201$ to $400$, etc). 
We recall that each run contains the outcomes from all the $9$ measurement steps in each execution. The row and column indexes denote, respectively, the number of time windows whose data are used to train and test the \ac{ml} model. Finally, the reported accuracy values are calculated by using the outcome distributions computed at all the measurement steps $k=1,\ldots,9$.
\\}
\label{tab:walkerSlow}
\centering
\includegraphics[width=0.94\linewidth]{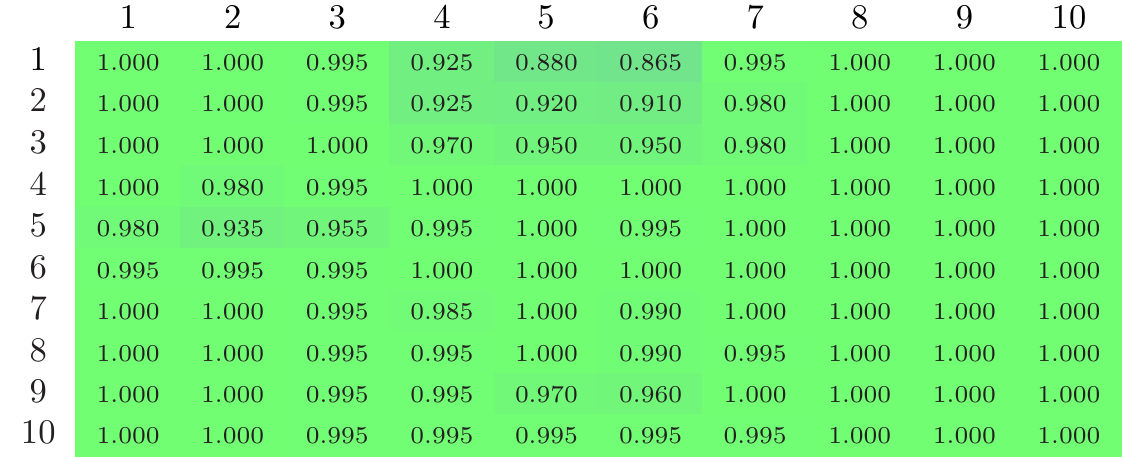}
\end{table*}
All the obtained results -- summarized in \cref{tab:walkerSlow} -- point out the following peculiar feature. Unsurprisingly, the \ac{svm} reaches $100\%$ of accuracy in the time window used for the training of the \ac{ml} model (corresponding to the diagonal of the table), 
and then, in proximity of the time windows on the diagonal, the accuracy decreases monotonically. This corresponds to the intuition that the machine-related noise fingerprint ``fade'' with time, 
due to the evidence -- discussed in the previous section -- that the noise fingerprint of the IBM quantum devices exhibits a quite prominent time-dependence. However, surprisingly, we observe that the accuracy returns to $100\%$ for time windows of runs 
far from the training one. We conjecture that this counter-intuitive phenomenon may be due either to the periodic calibration of the machines or to the slowdown induced by the fair-share queue. The latter, indeed, may be also observed in the last part of the \slow{} dataset in Fig.\,\ref{fig:walkerSlow}, and is supported by the evidence that, if we restrict the experiment 
to the runs from $1$ to $1\,000$ (i.e., the range where the execution times of the tested machines are more homogeneous as shown in Fig.\,\ref{fig:walkerSlow}), the resulting accuracy values decrease with time.

The general result that can be deduced from the experiments of the robustness analysis is that, by training our \ac{ml} model on just $200$ runs (corresponding to the diagonal time windows of the table), we are able to identify the device-related noise fingerprint with high accuracy for all the 1\,800 remaining ones. In this regard, it is worth noting that, between the training samples and the last test ones, there is up to a week in real-time execution (as one can see in Fig.\,\ref{fig:walkerSlow}). This means that we can consider our classifier to be fairly robust in time, despite the changes in the environment and calibration of the machines that might occur even at time-scales of weeks.

\section{Conclusions}

In this work we prove the existence of a noise fingerprint -- also admitting a clear time-dependent profile -- in the tested IBM quantum machines, which are just a particular class of NISQ devices. We have also demonstrate that such noise fingerprints can be exploited to reliably distinguish the machines by means of \ac{svm} models. As general results, our experiments confirm that (i) all the analysed quantum devices exhibit a clear machine-related noise fingerprint that is \emph{robust}, in the sense that the fingerprint is highly predictable over time in windows of consecutive runs; (ii) the noise fingerprint has also a \emph{time dependence}, namely it changes over time and after few hours
becomes different enough to be distinguished from fingerprint in the past; (iii) in each quantum device, sequences of measurement outcome distributions are required for the accurate learning of the corresponding noise fingerprint. One may conjecture that possible reason behind the latter aspect may be that the noisy dynamics in the IBM machines can be non-Markovian due to the presence of time-correlations among consecutive samples of the noise field. However, it is worth observing that the \acp{svm} we successfully used in this work are memory-less \ac{ml} models, which thus ignores possible temporal relations across the measurement steps. Therefore, the gathered data and the adopted \ac{ml} models are not indicated to validate any hypothesis on non-Markovianity. These aspects, deserving further investigations, will be addressed in another contribution in which memory-less \ac{ml} models will be compared with other \ac{ml} architectures processing time series data with variable memory length. In conclusion, despite the microscopic reasons for the existence of a machine-related noise fingerprint are still unknown (indeed, the IBM machines are partly inaccessible), we can now affirm that one can reliably leverage such noise profiles to distinguish, and possibly in the future characterize, different NISQ quantum devices.

As an outlook, learning the noise fingerprint of quantum devices from time-ordered measurements of testbed quantum circuits is expected to open the way, in the next future, to many other experiments and ideas. The proposed methodology, indeed, may be applied not only to IBM quantum machines, but even to a larger class of quantum devices, both in commercial or laboratory scenarios. In all of them, classification \ac{ml} model, exploiting the presence of intrinsic noise sources that give rise to an identifiable noise fingerprint in the devices, may be employed to predict on which machine, and at which time, a given quantum circuit or algorithm was executed. Moreover, our procedures could be adopted to predict if and when the noise fingerprint of a specific quantum device changes over time, e.g., due to calibration actions. Such a knowledge will help in mitigating (time-varying) errors occurring in the computation and, possibly, performing ad hoc error corrections.    

\paragraph{Data and code availability} 

The datasets and the source codes for the \ac{ml} experiments are available on \emph{GitHub} at the following address: \\
{\fontsize{8.7pt}{20pt}\selectfont\url{https://github.com/trianam/learningQuantumNoiseFingerprint}} .

\paragraph{Acknowledgments}

This work was financially supported from Fondazione CR Firenze through the project QUANTUM-AI, from University of Florence through the project Q-CODYCES, and from the European Union’s Horizon 2020 research and innovation programme under FET-OPEN Grant Agreement No.\,828946 (PATHOS). We acknowledge the access to advanced services provided by the IBM Quantum Researchers Program.

\bibliography{quantum,ml}


\begin{thebibliography}{40}
\ifx \bisbn   \undefined \def \bisbn  #1{ISBN #1}\fi
\ifx \binits  \undefined \def \binits#1{#1}\fi
\ifx \bauthor  \undefined \def \bauthor#1{#1}\fi
\ifx \batitle  \undefined \def \batitle#1{#1}\fi
\ifx \bjtitle  \undefined \def \bjtitle#1{#1}\fi
\ifx \bvolume  \undefined \def \bvolume#1{\textbf{#1}}\fi
\ifx \byear  \undefined \def \byear#1{#1}\fi
\ifx \bissue  \undefined \def \bissue#1{#1}\fi
\ifx \bfpage  \undefined \def \bfpage#1{#1}\fi
\ifx \blpage  \undefined \def \blpage #1{#1}\fi
\ifx \burl  \undefined \def \burl#1{\textsf{#1}}\fi
\ifx \doiurl  \undefined \def \doiurl#1{\url{https://doi.org/#1}}\fi
\ifx \betal  \undefined \def \betal{\textit{et al.}}\fi
\ifx \binstitute  \undefined \def \binstitute#1{#1}\fi
\ifx \binstitutionaled  \undefined \def \binstitutionaled#1{#1}\fi
\ifx \bctitle  \undefined \def \bctitle#1{#1}\fi
\ifx \beditor  \undefined \def \beditor#1{#1}\fi
\ifx \bpublisher  \undefined \def \bpublisher#1{#1}\fi
\ifx \bbtitle  \undefined \def \bbtitle#1{#1}\fi
\ifx \bedition  \undefined \def \bedition#1{#1}\fi
\ifx \bseriesno  \undefined \def \bseriesno#1{#1}\fi
\ifx \blocation  \undefined \def \blocation#1{#1}\fi
\ifx \bsertitle  \undefined \def \bsertitle#1{#1}\fi
\ifx \bsnm \undefined \def \bsnm#1{#1}\fi
\ifx \bsuffix \undefined \def \bsuffix#1{#1}\fi
\ifx \bparticle \undefined \def \bparticle#1{#1}\fi
\ifx \barticle \undefined \def \barticle#1{#1}\fi
\bibcommenthead
\ifx \bconfdate \undefined \def \bconfdate #1{#1}\fi
\ifx \botherref \undefined \def \botherref #1{#1}\fi
\ifx \url \undefined \def \url#1{\textsf{#1}}\fi
\ifx \bchapter \undefined \def \bchapter#1{#1}\fi
\ifx \bbook \undefined \def \bbook#1{#1}\fi
\ifx \bcomment \undefined \def \bcomment#1{#1}\fi
\ifx \oauthor \undefined \def \oauthor#1{#1}\fi
\ifx \citeauthoryear \undefined \def \citeauthoryear#1{#1}\fi
\ifx \endbibitem  \undefined \def \endbibitem {}\fi
\ifx \bconflocation  \undefined \def \bconflocation#1{#1}\fi
\ifx \arxivurl  \undefined \def \arxivurl#1{\textsf{#1}}\fi
\csname PreBibitemsHook\endcsname

\bibitem{preskill2018quantum}
\begin{barticle}
\bauthor{\bsnm{Preskill}, \binits{J.}}:
\batitle{Quantum {C}omputing in the {NISQ} era and beyond}.
\bjtitle{Quantum}
\bvolume{2},
\bfpage{79}
(\byear{2018})
\end{barticle}
\endbibitem

\bibitem{deutsch2020harnessing}
\begin{barticle}
\bauthor{\bsnm{Deutsch}, \binits{I.H.}}:
\batitle{Harnessing the power of the second quantum revolution}.
\bjtitle{PRX Quantum}
\bvolume{1}(\bissue{2}),
\bfpage{020101}
(\byear{2020})
\end{barticle}
\endbibitem

\bibitem{BhartiArxiv2021}
\begin{botherref}
\oauthor{\bsnm{Bharti}, \binits{K.}},
\oauthor{\bsnm{Cervera-Lierta}, \binits{A.}},
\oauthor{\bsnm{Kyaw}, \binits{T.H.}},
\oauthor{\bsnm{Haug}, \binits{T.}},
\oauthor{\bsnm{Alperin-Lea}, \binits{S.}},
\oauthor{\bsnm{Anand}, \binits{A.}},
\oauthor{\bsnm{Degroote}, \binits{M.}},
\oauthor{\bsnm{Heimonen}, \binits{H.}},
\oauthor{\bsnm{Kottmann}, \binits{J.S.}},
\oauthor{\bsnm{Menke}, \binits{T.}},
\oauthor{\bsnm{Mok}, \binits{W.-K.}},
\oauthor{\bsnm{Sim}, \binits{S.}},
\oauthor{\bsnm{Kwek}, \binits{L.-C.}},
\oauthor{\bsnm{Aspuru-Guzik}, \binits{A.}}:
Noisy intermediate-scale quantum {(NISQ)} algorithms.
arXiv preprint arXiv:2101.08448
(2021)
\end{botherref}
\endbibitem

\bibitem{DegenRMP2017}
\begin{barticle}
\bauthor{\bsnm{Degen}, \binits{C.L.}},
\bauthor{\bsnm{Reinhard}, \binits{F.}},
\bauthor{\bsnm{Cappellaro}, \binits{P.}}:
\batitle{Quantum sensing}.
\bjtitle{Reviews of Modern Physics}
\bvolume{89}(\bissue{3}),
\bfpage{035002}
(\byear{2017})
\end{barticle}
\endbibitem

\bibitem{SzankowskiJPCM2017}
\begin{barticle}
\bauthor{\bsnm{Sza{\'n}kowski}, \binits{P.}},
\bauthor{\bsnm{Ramon}, \binits{G.}},
\bauthor{\bsnm{Krzywda}, \binits{J.}},
\bauthor{\bsnm{Kwiatkowski}, \binits{D.}}, \betal:
\batitle{Environmental noise spectroscopy with qubits subjected to dynamical
  decoupling}.
\bjtitle{Journal of Physics: Condensed Matter}
\bvolume{29}(\bissue{33}),
\bfpage{333001}
(\byear{2017})
\end{barticle}
\endbibitem

\bibitem{DoNJP2019}
\begin{barticle}
\bauthor{\bsnm{Do}, \binits{H.-V.}},
\bauthor{\bsnm{Lovecchio}, \binits{C.}},
\bauthor{\bsnm{Mastroserio}, \binits{I.}},
\bauthor{\bsnm{Fabbri}, \binits{N.}},
\bauthor{\bsnm{Cataliotti}, \binits{F.S.}},
\bauthor{\bsnm{Gherardini}, \binits{S.}},
\bauthor{\bsnm{M{\"u}ller}, \binits{M.M.}},
\bauthor{\bsnm{Dalla~Pozza}, \binits{N.}},
\bauthor{\bsnm{Caruso}, \binits{F.}}:
\batitle{Experimental proof of quantum {Z}eno-assisted noise sensing}.
\bjtitle{New Journal of Physics}
\bvolume{21}(\bissue{11}),
\bfpage{113056}
(\byear{2019})
\end{barticle}
\endbibitem

\bibitem{MuellerPLA2020}
\begin{barticle}
\bauthor{\bsnm{M{\"u}ller}, \binits{M.M.}},
\bauthor{\bsnm{Gherardini}, \binits{S.}},
\bauthor{\bsnm{Dalla~Pozza}, \binits{N.}},
\bauthor{\bsnm{Caruso}, \binits{F.}}:
\batitle{Noise sensing via stochastic quantum {Z}eno}.
\bjtitle{Physics Letters A}
\bvolume{384}(\bissue{13}),
\bfpage{126244}
(\byear{2020})
\end{barticle}
\endbibitem

\bibitem{wise2021using}
\begin{barticle}
\bauthor{\bsnm{Wise}, \binits{D.F.}},
\bauthor{\bsnm{Morton}, \binits{J.J.L.}},
\bauthor{\bsnm{Dhomkar}, \binits{S.}}:
\batitle{Using deep learning to understand and mitigate the qubit noise
  environment}.
\bjtitle{PRX Quantum}
\bvolume{2},
\bfpage{010316}
(\byear{2021})
\end{barticle}
\endbibitem

\bibitem{ColeNanotech2009}
\begin{barticle}
\bauthor{\bsnm{Cole}, \binits{J.H.}},
\bauthor{\bsnm{Hollenberg}, \binits{L.C.}}:
\batitle{Scanning quantum decoherence microscopy}.
\bjtitle{Nanotechnology}
\bvolume{20}(\bissue{49}),
\bfpage{495401}
(\byear{2009})
\end{barticle}
\endbibitem

\bibitem{BylanderNatPhys2011}
\begin{barticle}
\bauthor{\bsnm{Bylander}, \binits{J.}},
\bauthor{\bsnm{Gustavsson}, \binits{S.}},
\bauthor{\bsnm{Yan}, \binits{F.}},
\bauthor{\bsnm{Yoshihara}, \binits{F.}},
\bauthor{\bsnm{Harrabi}, \binits{K.}},
\bauthor{\bsnm{Fitch}, \binits{G.}},
\bauthor{\bsnm{Cory}, \binits{D.G.}},
\bauthor{\bsnm{Nakamura}, \binits{Y.}},
\bauthor{\bsnm{Tsai}, \binits{J.-S.}},
\bauthor{\bsnm{Oliver}, \binits{W.D.}}:
\batitle{Noise spectroscopy through dynamical decoupling with a superconducting
  flux qubit}.
\bjtitle{Nature Physics}
\bvolume{7}(\bissue{7}),
\bfpage{565}--\blpage{570}
(\byear{2011})
\end{barticle}
\endbibitem

\bibitem{AlvarezPRL2011}
\begin{barticle}
\bauthor{\bsnm{{\'A}lvarez}, \binits{G.A.}},
\bauthor{\bsnm{Suter}, \binits{D.}}:
\batitle{Measuring the spectrum of colored noise by dynamical decoupling}.
\bjtitle{Physical Review Letters}
\bvolume{107}(\bissue{23}),
\bfpage{230501}
(\byear{2011})
\end{barticle}
\endbibitem

\bibitem{YugePRL2011}
\begin{barticle}
\bauthor{\bsnm{Yuge}, \binits{T.}},
\bauthor{\bsnm{Sasaki}, \binits{S.}},
\bauthor{\bsnm{Hirayama}, \binits{Y.}}:
\batitle{Measurement of the noise spectrum using a multiple-pulse sequence}.
\bjtitle{Physical Review Letters}
\bvolume{107}(\bissue{17}),
\bfpage{170504}
(\byear{2011})
\end{barticle}
\endbibitem

\bibitem{Paz-SilvaPRL2014}
\begin{barticle}
\bauthor{\bsnm{Paz-Silva}, \binits{G.A.}},
\bauthor{\bsnm{Viola}, \binits{L.}}:
\batitle{General transfer-function approach to noise filtering in open-loop
  quantum control}.
\bjtitle{Physical Review Letters}
\bvolume{113}(\bissue{25}),
\bfpage{250501}
(\byear{2014})
\end{barticle}
\endbibitem

\bibitem{NorrisPRL2016}
\begin{barticle}
\bauthor{\bsnm{Norris}, \binits{L.M.}},
\bauthor{\bsnm{Paz-Silva}, \binits{G.A.}},
\bauthor{\bsnm{Viola}, \binits{L.}}:
\batitle{Qubit noise spectroscopy for non-{G}aussian dephasing environments}.
\bjtitle{Physical Review Letters}
\bvolume{116}(\bissue{15}),
\bfpage{150503}
(\byear{2016})
\end{barticle}
\endbibitem

\bibitem{FreyNatComm2017}
\begin{barticle}
\bauthor{\bsnm{Frey}, \binits{V.M.}},
\bauthor{\bsnm{Mavadia}, \binits{S.}},
\bauthor{\bsnm{Norris}, \binits{L.M.}},
\bauthor{\bparticle{de} \bsnm{Ferranti}, \binits{W.}},
\bauthor{\bsnm{Lucarelli}, \binits{D.}},
\bauthor{\bsnm{L.}, \binits{V.}},
\bauthor{\bsnm{Biercuk}, \binits{M.J.}}:
\batitle{Application of optimal band-limited control protocols to quantum noise
  sensing}.
\bjtitle{Nat. Commun.}
\bvolume{8},
\bfpage{2189}
(\byear{2017})
\end{barticle}
\endbibitem

\bibitem{MuellerSciRep2018}
\begin{barticle}
\bauthor{\bsnm{M{\"u}ller}, \binits{M.M.}},
\bauthor{\bsnm{Gherardini}, \binits{S.}},
\bauthor{\bsnm{Caruso}, \binits{F.}}:
\batitle{Noise-robust quantum sensing via optimal multi-probe spectroscopy}.
\bjtitle{Scientific Reports}
\bvolume{8}(\bissue{1}),
\bfpage{1}--\blpage{17}
(\byear{2018})
\end{barticle}
\endbibitem

\bibitem{HernandezPRB2018}
\begin{barticle}
\bauthor{\bsnm{Hern{\'a}ndez-G{\'o}mez}, \binits{S.}},
\bauthor{\bsnm{Poggiali}, \binits{F.}},
\bauthor{\bsnm{Cappellaro}, \binits{P.}},
\bauthor{\bsnm{Fabbri}, \binits{N.}}:
\batitle{Noise spectroscopy of a quantum-classical environment with a diamond
  qubit}.
\bjtitle{Physical Review B}
\bvolume{98}(\bissue{21}),
\bfpage{214307}
(\byear{2018})
\end{barticle}
\endbibitem

\bibitem{GomezFrontiers2021}
\begin{barticle}
\bauthor{\bsnm{Hern\'andez-G\'omez}, \binits{S.}},
\bauthor{\bsnm{Fabbri}, \binits{N.}}:
\batitle{Quantum control for nanoscale spectroscopy with diamond
  nitrogen-vacancy centers: A short review}.
\bjtitle{Frontiers in Physics}
\bvolume{8},
\bfpage{610868}
(\byear{2021})
\end{barticle}
\endbibitem

\bibitem{devoret2004superconducting}
\begin{botherref}
\oauthor{\bsnm{Devoret}, \binits{M.H.}},
\oauthor{\bsnm{Wallraff}, \binits{A.}},
\oauthor{\bsnm{Martinis}, \binits{J.M.}}:
Superconducting qubits: A short review.
arXiv preprint cond-mat/0411174
(2004)
\end{botherref}
\endbibitem

\bibitem{clarke2008superconducting}
\begin{barticle}
\bauthor{\bsnm{Clarke}, \binits{J.}},
\bauthor{\bsnm{Wilhelm}, \binits{F.K.}}:
\batitle{Superconducting quantum bits}.
\bjtitle{Nature}
\bvolume{453}(\bissue{7198}),
\bfpage{1031}--\blpage{1042}
(\byear{2008})
\end{barticle}
\endbibitem

\bibitem{wineland2003quantum}
\begin{barticle}
\bauthor{\bsnm{Wineland}, \binits{D.J.}},
\bauthor{\bsnm{Barrett}, \binits{M.}},
\bauthor{\bsnm{Britton}, \binits{J.}},
\bauthor{\bsnm{Chiaverini}, \binits{J.}},
\bauthor{\bsnm{DeMarco}, \binits{B.}},
\bauthor{\bsnm{Itano}, \binits{W.M.}},
\bauthor{\bsnm{Jelenkovi{\'c}}, \binits{B.}},
\bauthor{\bsnm{Langer}, \binits{C.}},
\bauthor{\bsnm{Leibfried}, \binits{D.}},
\bauthor{\bsnm{Meyer}, \binits{V.}}, \betal:
\batitle{Quantum information processing with trapped ions}.
\bjtitle{Philosophical Transactions of the Royal Society of London. Series A:
  Mathematical, Physical and Engineering Sciences}
\bvolume{361}(\bissue{1808}),
\bfpage{1349}--\blpage{1361}
(\byear{2003})
\end{barticle}
\endbibitem

\bibitem{spring2013boson}
\begin{barticle}
\bauthor{\bsnm{Spring}, \binits{J.B.}},
\bauthor{\bsnm{Metcalf}, \binits{B.J.}},
\bauthor{\bsnm{Humphreys}, \binits{P.C.}},
\bauthor{\bsnm{Kolthammer}, \binits{W.S.}},
\bauthor{\bsnm{Jin}, \binits{X.-M.}},
\bauthor{\bsnm{Barbieri}, \binits{M.}},
\bauthor{\bsnm{Datta}, \binits{A.}},
\bauthor{\bsnm{Thomas-Peter}, \binits{N.}},
\bauthor{\bsnm{Langford}, \binits{N.K.}},
\bauthor{\bsnm{Kundys}, \binits{D.}}, \betal:
\batitle{Boson sampling on a photonic chip}.
\bjtitle{Science}
\bvolume{339}(\bissue{6121}),
\bfpage{798}--\blpage{801}
(\byear{2013})
\end{barticle}
\endbibitem

\bibitem{metcalf2014quantum}
\begin{barticle}
\bauthor{\bsnm{Metcalf}, \binits{B.J.}},
\bauthor{\bsnm{Spring}, \binits{J.B.}},
\bauthor{\bsnm{Humphreys}, \binits{P.C.}},
\bauthor{\bsnm{Thomas-Peter}, \binits{N.}},
\bauthor{\bsnm{Barbieri}, \binits{M.}},
\bauthor{\bsnm{Kolthammer}, \binits{W.S.}},
\bauthor{\bsnm{Jin}, \binits{X.-M.}},
\bauthor{\bsnm{Langford}, \binits{N.K.}},
\bauthor{\bsnm{Kundys}, \binits{D.}},
\bauthor{\bsnm{Gates}, \binits{J.C.}}, \betal:
\batitle{Quantum teleportation on a photonic chip}.
\bjtitle{Nature photonics}
\bvolume{8}(\bissue{10}),
\bfpage{770}--\blpage{774}
(\byear{2014})
\end{barticle}
\endbibitem

\bibitem{freedman2003topological}
\begin{barticle}
\bauthor{\bsnm{Freedman}, \binits{M.}},
\bauthor{\bsnm{Kitaev}, \binits{A.}},
\bauthor{\bsnm{Larsen}, \binits{M.}},
\bauthor{\bsnm{Wang}, \binits{Z.}}:
\batitle{Topological quantum computation}.
\bjtitle{Bulletin of the American Mathematical Society}
\bvolume{40}(\bissue{1}),
\bfpage{31}--\blpage{38}
(\byear{2003})
\end{barticle}
\endbibitem

\bibitem{BishopPRML2006}
\begin{bbook}
\bauthor{\bsnm{Bishop}, \binits{C.M.}}:
\bbtitle{Pattern Recognition and Machine Learning},
\bedition{1}st edn.
\bpublisher{Springer},
\blocation{New York}
(\byear{2006})
\end{bbook}
\endbibitem

\bibitem{HastieESL2009}
\begin{bbook}
\bauthor{\bsnm{Hastie}, \binits{T.}},
\bauthor{\bsnm{Tibshirani}, \binits{R.}},
\bauthor{\bsnm{Friedman}, \binits{J.}}:
\bbtitle{The Elements of Statistical Learning: Data Mining, Inference, and
  Prediction},
\bedition{2}nd edn.
\bpublisher{Springer},
\blocation{New York}
(\byear{2009})
\end{bbook}
\endbibitem

\bibitem{YoussryArXiv}
\begin{barticle}
\bauthor{\bsnm{Youssry}, \binits{A.}},
\bauthor{\bsnm{Paz-Silva}, \binits{G.A.}},
\bauthor{\bsnm{Ferrie}, \binits{C.}}:
\batitle{Beyond {Q}uantum {N}oise {S}pectroscopy: modelling and mitigating
  noise with quantum feature engineering}.
\bjtitle{npj Quantum Information}
\bvolume{6},
\bfpage{95}
(\byear{2020})
\end{barticle}
\endbibitem

\bibitem{luchnikov2020machine}
\begin{barticle}
\bauthor{\bsnm{Luchnikov}, \binits{I.}},
\bauthor{\bsnm{Vintskevich}, \binits{S.}},
\bauthor{\bsnm{Grigoriev}, \binits{D.}},
\bauthor{\bsnm{Filippov}, \binits{S.}}:
\batitle{Machine learning non-{M}arkovian quantum dynamics}.
\bjtitle{Physical Review Letters}
\bvolume{124}(\bissue{14}),
\bfpage{140502}
(\byear{2020})
\end{barticle}
\endbibitem

\bibitem{fanchini2020estimating}
\begin{barticle}
\bauthor{\bsnm{Fanchini}, \binits{F.F.}},
\bauthor{\bsnm{Karpat}, \binits{G.b.u.}},
\bauthor{\bsnm{Rossatto}, \binits{D.Z.}},
\bauthor{\bsnm{Norambuena}, \binits{A.}},
\bauthor{\bsnm{Coto}, \binits{R.}}:
\batitle{Estimating the degree of non-{M}arkovianity using machine learning}.
\bjtitle{Physical Review A}
\bvolume{103},
\bfpage{022425}
(\byear{2021})
\end{barticle}
\endbibitem

\bibitem{niu2019learning}
\begin{botherref}
\oauthor{\bsnm{Niu}, \binits{M.Y.}},
\oauthor{\bsnm{Smelyanskyi}, \binits{V.}},
\oauthor{\bsnm{Klimov}, \binits{P.}},
\oauthor{\bsnm{Boixo}, \binits{S.}},
\oauthor{\bsnm{Barends}, \binits{R.}},
\oauthor{\bsnm{Kelly}, \binits{J.}},
\oauthor{\bsnm{Chen}, \binits{Y.}},
\oauthor{\bsnm{Arya}, \binits{K.}},
\oauthor{\bsnm{Burkett}, \binits{B.}},
\oauthor{\bsnm{Bacon}, \binits{D.}}, et al.:
Learning non-{M}arkovian quantum noise from {M}oir\'e-enhanced swap
  spectroscopy with deep evolutionary algorithm.
arXiv preprint arXiv:1912.04368
(2019)
\end{botherref}
\endbibitem

\bibitem{HarperNatPhys2020}
\begin{botherref}
\oauthor{\bsnm{Harper}, \binits{R.}},
\oauthor{\bsnm{Flammia}, \binits{S.T.}},
\oauthor{\bsnm{Wallman}, \binits{J.J.}}:
Efficient learning of quantum noise.
Nature Physics,
1--5
(2020)
\end{botherref}
\endbibitem

\bibitem{MartinaArXiv2021}
\begin{botherref}
\oauthor{\bsnm{Martina}, \binits{S.}},
\oauthor{\bsnm{Gherardini}, \binits{S.}},
\oauthor{\bsnm{Caruso}, \binits{F.}}:
Machine learning approach for quantum non-{M}arkovian noise classification.
arXiv preprint arXiv:2101.03221
(2021)
\end{botherref}
\endbibitem

\bibitem{ibmq}
\begin{botherref}
\url{https://quantum-computing.ibm.com/}
(Visited on 2021)
\end{botherref}
\endbibitem

\bibitem{eisert2020quantum}
\begin{barticle}
\bauthor{\bsnm{Eisert}, \binits{J.}},
\bauthor{\bsnm{Hangleiter}, \binits{D.}},
\bauthor{\bsnm{Walk}, \binits{N.}},
\bauthor{\bsnm{Roth}, \binits{I.}},
\bauthor{\bsnm{Markham}, \binits{D.}},
\bauthor{\bsnm{Parekh}, \binits{R.}},
\bauthor{\bsnm{Chabaud}, \binits{U.}},
\bauthor{\bsnm{Kashefi}, \binits{E.}}:
\batitle{Quantum certification and benchmarking}.
\bjtitle{Nature Reviews Physics}
\bvolume{2}(\bissue{7}),
\bfpage{382}--\blpage{390}
(\byear{2020})
\end{barticle}
\endbibitem

\bibitem{qiskit}
\begin{botherref}
\oauthor{\bparticle{et} \bsnm{al.}, \binits{H.A.}}:
Qiskit: An Open-source Framework for Quantum Computing
(2019)
\end{botherref}
\endbibitem

\bibitem{cross2019validating}
\begin{barticle}
\bauthor{\bsnm{Cross}, \binits{A.W.}},
\bauthor{\bsnm{Bishop}, \binits{L.S.}},
\bauthor{\bsnm{Sheldon}, \binits{S.}},
\bauthor{\bsnm{Nation}, \binits{P.D.}},
\bauthor{\bsnm{Gambetta}, \binits{J.M.}}:
\batitle{Validating quantum computers using randomized model circuits}.
\bjtitle{Physical Review A}
\bvolume{100}(\bissue{3}),
\bfpage{032328}
(\byear{2019})
\end{barticle}
\endbibitem

\bibitem{FischerPRL2001}
\begin{barticle}
\bauthor{\bsnm{Fischer}, \binits{M.C.}},
\bauthor{\bsnm{Guti\'errez-Medina}, \binits{B.}},
\bauthor{\bsnm{Raizen}, \binits{M.G.}}:
\batitle{Observation of the quantum {Z}eno and {A}nti-{Z}eno effects in an
  unstable system}.
\bjtitle{Phys. Rev. Lett.}
\bvolume{87},
\bfpage{040402}
(\byear{2001})
\end{barticle}
\endbibitem

\bibitem{SchaferNatComm2014}
\begin{barticle}
\bauthor{\bsnm{Sch\"{a}fer}, \binits{F.}},
\bauthor{\bsnm{Herrera}, \binits{I.}},
\bauthor{\bsnm{Cherukattil}, \binits{S.}},
\bauthor{\bsnm{Lovecchio}, \binits{C.}},
\bauthor{\bsnm{Cataliotti}, \binits{F.S.}},
\bauthor{\bsnm{Caruso}, \binits{F.}},
\bauthor{\bsnm{Smerzi}, \binits{A.}}:
\batitle{Experimental realization of quantum zeno dynamics}.
\bjtitle{Nat. Commun.}
\bvolume{5},
\bfpage{3194}
(\byear{2014})
\end{barticle}
\endbibitem

\bibitem{GherardiniQST2017}
\begin{barticle}
\bauthor{\bsnm{Gherardini}, \binits{S.}},
\bauthor{\bsnm{Lovecchio}, \binits{C.}},
\bauthor{\bsnm{M\"{u}ller}, \binits{M.M.}},
\bauthor{\bsnm{Lombardi}, \binits{P.}},
\bauthor{\bsnm{Caruso}, \binits{F.}},
\bauthor{\bsnm{Cataliotti}, \binits{F.S.}}:
\batitle{Ergodicity in randomly perturbed quantum systems}.
\bjtitle{Quantum Sci. Technol.}
\bvolume{2},
\bfpage{015007}
(\byear{2017})
\end{barticle}
\endbibitem

\bibitem{VirziArXiv2021}
\begin{botherref}
\oauthor{\bsnm{Virz\'i}, \binits{S.}},
\oauthor{\bsnm{Avella}, \binits{A.}},
\oauthor{\bsnm{Piacentini}, \binits{F.}},
\oauthor{\bsnm{Gramegna}, \binits{M.}},
\oauthor{\bsnm{Opatrny}, \binits{T.}},
\oauthor{\bsnm{Kofman}, \binits{A.}},
\oauthor{\bsnm{Kurizki}, \binits{G.}},
\oauthor{\bsnm{Gherardini}, \binits{S.}},
\oauthor{\bsnm{Caruso}, \binits{F.}},
\oauthor{\bsnm{Degiovanni}, \binits{I.P.}},
\oauthor{\bsnm{Genovese}, \binits{M.}}:
Quantum {Z}eno and {A}nti-{Z}eno probes of noise correlations in photon
  polarisation.
arXiv preprint arXiv:2103.03698
(2021)
\end{botherref}
\endbibitem

\end{thebibliography}

\end{document}